\documentclass[]{aa}  
\usepackage{xcolor}
\usepackage{txfonts}
\usepackage{hyperref}
\usepackage{amssymb}
\usepackage{graphicx}
\usepackage{amsmath,bm}
\usepackage{siunitx}
\usepackage{pict2e}
\usepackage{multirow}
\usepackage{dsfont}
\usepackage{appendix}

\def\ii{{\rm i}}

\def\br{\mbox{\boldmath$r$}}
\def\bxi{\mbox{\boldmath$\xi$}}
\begin{document} 
\title{Imaging individual active regions on the Sun's far side with improved helioseismic holography}
\author{
Dan Yang\inst{1}
\and Laurent Gizon\inst{1,2,3}
\and H\'el\`ene Barucq\inst{4}
}
\institute{Max-Planck-Institut f\"ur Sonnensystemforschung,  Justus-von-Liebig-Weg 3, 37077 G{\"o}ttingen,  Germany \\ \email{yangd@mps.mpg.de} 
\and
    Institut f\"ur Astrophysik, Georg-August-Universit{\"a}t G\"ottingen, Friedrich-Hund-Platz 1, 37077 G{\"o}ttingen, Germany   
       \and 
          Center for Space Science, NYUAD Institute, New York University Abu Dhabi, PO Box 129188, Abu Dhabi, UAE
            \and 
               Makutu, Inria, TotalEnergies, University of Pau, 64000 Pau, France
              }
              
\date{Received \today ;  Accepted <date>}
\abstract
{Helioseismic holography is a useful method to detect active regions on the Sun's far side and improve  space weather forecasts. 
}
{We aim to improve helioseismic holography by using a clear formulation of the problem, an accurate forward solver in the frequency domain, and a better understanding of the noise properties.
}
{Building on the work of  {Lindsey et al.}, we define the forward- and backward-propagated wave fields (ingression and egression) in terms of a Green's function. This Green's function is computed using an accurate forward solver in the frequency domain. 
We analyse overlapping segments of 31~hr of SDO/HMI dopplergrams, with a cadence of 24 hr. Phase shifts between the ingression and the egression are measured and averaged to detect active regions on the far side. }
{The phase maps are compared with direct EUV intensity maps from STEREO/EUVI. We confirm that medium-size active regions can be detected on the far side with high confidence. Their evolution (and possible emergence) can be monitored on a daily time scale. Seismic maps averaged over 3 days provide an active region detection rate as high as 75\% and a false discovery rate only as low as 7\%, for active regions with areas above one thousandth of an hemisphere.  {For a large part, these improvements can be attributed to the use of a complete Green's function (all skips) and to the use of all observations on the front side (full pupil).}
}
{Improved helioseismic holography enables the study of the evolution of medium-size active regions on the Sun's far side.  
}

\maketitle

\section{Introduction}
\label{sec.intro}
Detecting active regions on the Sun's far side is potentially of great importance for space-weather forecasts \citep[e.g.][]{ARG13,HILL18}.
Large active regions that emerge on the Sun's far side will rotate into Earth's view several days later; these may trigger CMEs that could damage satellites and spacecrafts and endanger astronauts  \citep[e.g.,][]{Lin11}. It is known that farside imaging can  significantly improve models of the solar wind \citep{ARG13}, which play an important role in  space-weather forecasts.

Two approaches have been considered so far to monitor the Sun's far side. The first solution consists of sending spacecrafts with direct views of the far side. The two STEREO spacecrafts (in trailing and leading orbits) provided images of the Sun's corona over the full range of S/C-Sun-Earth angles in the ecliptic plane \citep[e.g.,][]{Howard2008}. Solar Orbiter,  now in  operation, also provides direct images of the Sun from a sophisticated flight path out of the ecliptic, including magnetograms \citep[][]{MUL20,SOL20}. The other approach to image the Sun's far side is helioseismology, using high-cadence data from, e.g., the SDO spacecraft  or the GONG ground-based network. The concept of farside helioseismology was unequivocally proved to work  by \citet{LIN00b}. This approach is admittedly challenging, however it does not require the need to place additional  spacecrafts along non-trivial flight paths.

Acoustic waves  propagate horizontally and are trapped in the vertical direction -- they connect the Sun's near and far sides. Since acoustic waves travel faster in magnetized regions, they can inform us about the presence of active regions along their paths of propagation.  Two helioseismic techniques, helioseismic holography  \citep[e.g.,][]{LIN00b,BRA02,LIN17}  and time-distance helioseismology \citep[e.g.,][]{ZHA07, ILO09,ZHA19}  have been used to  detect active regions on the far side. In the present article, we restrict our attention to helioseismic holography.

Farside images computed using holography are  publicly available. These included  the JSOC/Stanford data set \footnote{\url{http://jsoc.stanford.edu/data/farside/}}and the NSO/GONG data set \footnote{ \url{https://farside.nso.edu/}}. 
These data sets were validated by comparison with the STEREO/EUVI observations  \citep{LIE12, Liewer2014,LIE17} and maps of the magnetic field \citep{GON07}. 
Both pipelines use a delta function (single ray) representation  of the Green's function.
While these farside maps have revealed  large active regions on the  far side, their quality is far from optimal. Several improvements have been proposed. 
For example, \cite{PRE10}  proposed to used WKB approximations to the  Green's functions, and to combine the  two-skip and three-skip seismic wave paths.
\citet{FEL19} used a machine learning algorithm  trained with Earth-side active regions to help detect small regions originally buried in noise \citep[also see][]{BRO21}. 

This work aims to improve holographic farside images by taking  advantage of recent theoretical and numerical advances in helioseismic imaging. In particular, we wish to work under the framework proposed by \citet{GIZ18} to define signal and noise and we wish to use an accurate and efficient forward solver to compute the finite-wavelength Green's functions  \citep{GIZ17}. We modify the  JSOC/Stanford farside imaging pipeline to include these updates. The resulting  maps are then compared to the  STEREO/EUVI observations when possible, and to other existing seismic maps.

\section{Helioseismic holography}
\label{sec.preliminaries}
The basic steps of seismic holography are rather simple:
1) numerically propagate the observed wavefield on the Sun's near side to target locations on the far side, 2) multiply the forward and backward propagated wavefields and measure a phase map, and then 3) subtract a reference measurement for the quiet Sun. These steps are described in detail in the following sections, and definitions are given for signal and noise.

\subsection{Wave equation}
\label{sec.holo_definition}
Following \citet{DEU84}, we consider the scalar variable 
\begin{equation}
\psi(\br,\omega) =  \rho^{1/2}(\br)  c^2(\br)  \nabla\cdot \bxi(\br,\omega), \label{eq.psi}
\end{equation} 
where  $\rho$  and $c$  are the density and sound speed  at  position $\br$, and $\bxi$ is the wave displacement vector at  $\br$ and angular frequency $\omega$. Ignoring terms involving gravity, this change of variable leads to a Helmholtz-like equation for $\psi$:
\begin{equation}
  - \left[\nabla^2 + k^2(\br,\omega)\right] \psi (\br,\omega) = S(\br,\omega), \label{eq.waveeq}
\end{equation}
where $k$ is the local wavenumber 
\begin{equation}
k^2  = \left( \omega^2 + 2\ii \omega \gamma \right) / c^2-  \rho^{1/2}  \nabla^2 \left( \rho^{-1/2} \right) . \label{eq.wavenumber}
\end{equation}
In the frequency range of interest ($2.5$--$4.5$~mHz), we take the damping rate to be 
$\gamma = \gamma_0 \left|{\omega}/{\omega_0}\right|^{5.77}$, where $\gamma_0/2\pi=4.29\ \mu\mathrm{Hz}$ and $\omega_0/2\pi=3\ \mathrm{mHz}$ \citep[][]{GIZ17}. 
The quantity $S$ in the above equation is a source term that represents wave excitation by turbulent convection.
In writing the above equations, we assumed the following Fourier convention: 
\begin{equation}
F(\omega) = \int f(t) e^{\mathrm{i} \omega t} \mathrm{d}t
\end{equation}
for any function of time $f(t)$ that decays fast enough at infinities.

\begin{figure}[!htb]
\begin{center}
\includegraphics[width=\linewidth]{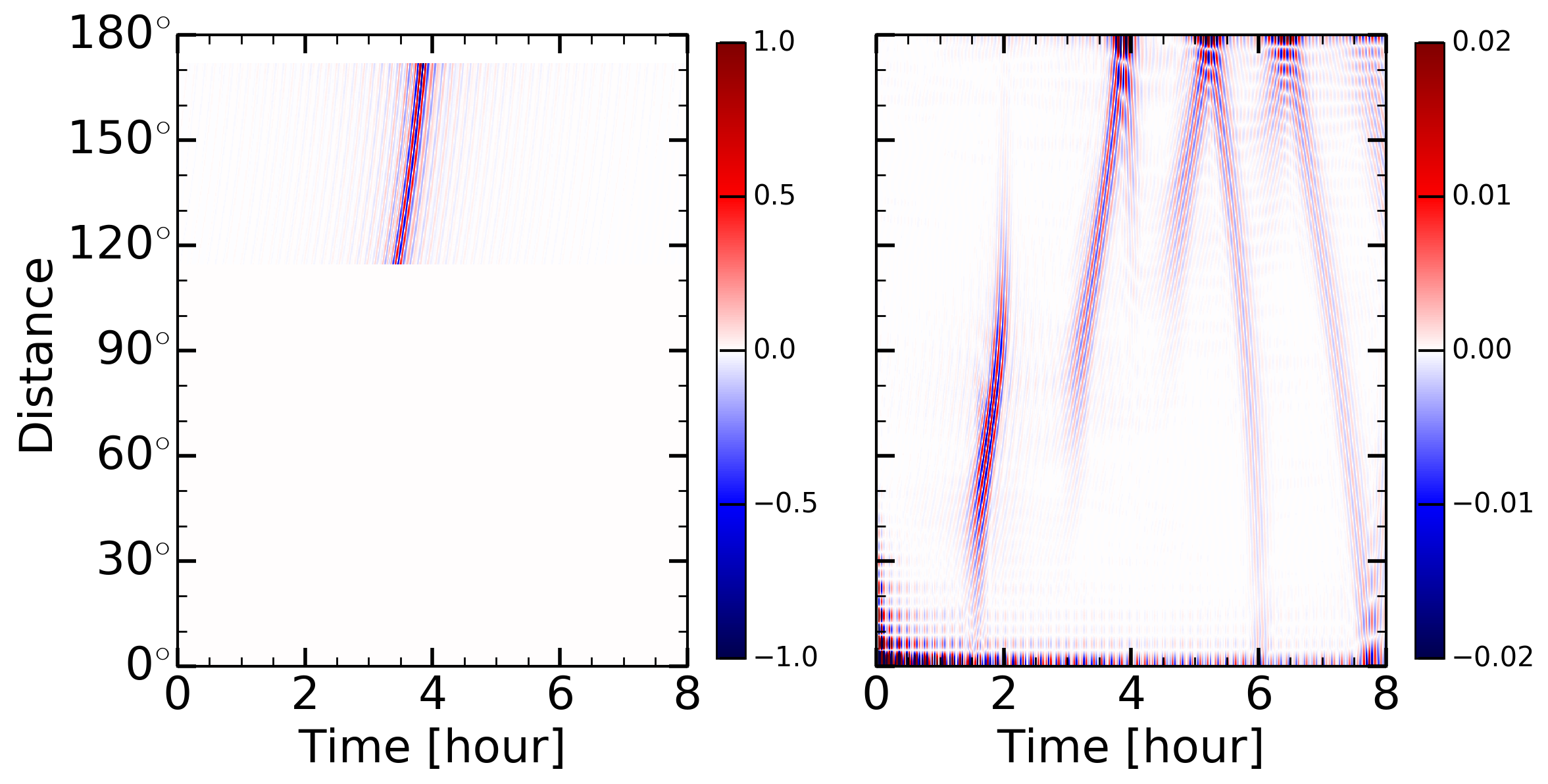}
\caption{
Comparison between two choices of Green's functions used in helioseismic holography, as functions of the angular distance $\Delta$ between the source and the receiver (both on the solar surface) and time. In both cases the frequency bandwidth is $2.5$--$4.5$~mHz. 
{\it Left panel}: Second-skip Green's function based on the formalism of \citet{LIN00a} (corrected ray theory) and used by the JSOC/Stanford farside imaging pipeline.
{Distances from  115$^{\circ}$ to 172$^{\circ}$ are used to map the far side within $50^\circ$ of the antipode to disk center.} 
{\it Right panel}: Green's function obtained by solving the scalar wave equation in the frequency domain (Eq.~\ref{eq.greens}). The solution is restricted to harmonic degrees in the range 5--45.}
\label{fig1}
\end{center}
\end{figure}

\begin{figure*}[!htb]
\begin{center}
\includegraphics[width=\linewidth]{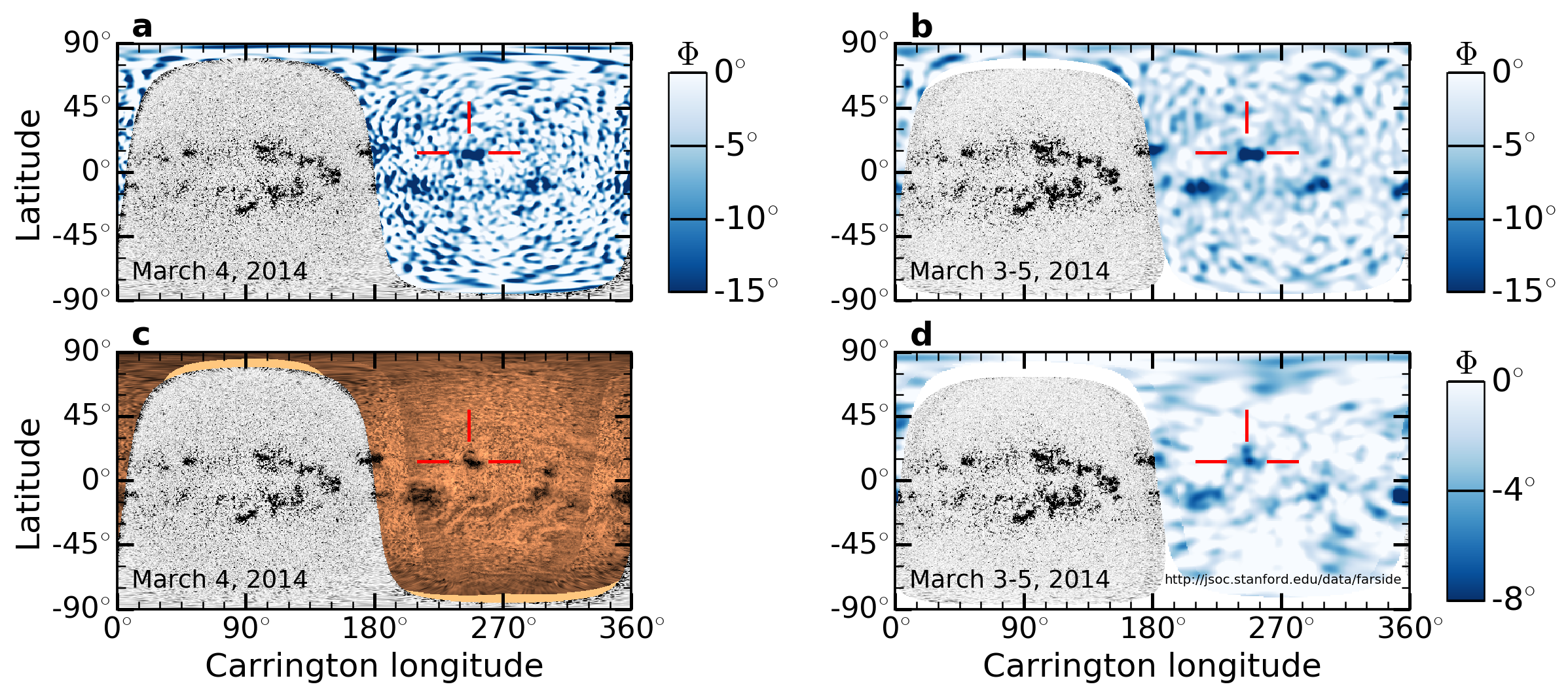} 
\caption{
Composite maps of the solar surface in the Carrington frame. In all panels, the near side  (gray shades) shows the SDO/HMI unsigned line-of-sight magnetic field at 0:00~TAI on March 4, 2014. {\it Panel (a)}: Farside map using improved helioseismic holography ($\Phi$, blue shades) applied to 31~hr of SDO/HMI Dopplergrams centered on March 4, 2014.
The red lines indicate the location of active region NOAA 12007 when it emerges $\sim\!8$ days later on the near side. The longitude is that of Carrington rotation~2147.
{\it Panel (b)}: Same as panel (a) but for a 3-day average over March 3--5. A low pass spatial filter ($l < 40$) was applied. 
{\it Panel (c)}: Far side seen by STEREO/EUVI at 304 \si{\angstrom}. 
{\it Panel (d)}: Farside map from the standard JSOC/Stanford helioseismic pipeline (blue shades) using $3$ days of data. }  
\label{fig2}
\end{center}
\end{figure*}

\subsection{Wave propagators}
\label{sec.egreesion}
The fundamental idea is to detect wave scattering by sunspots and other magnetic regions on the far side via the phase shift between the forward-propagated and the back-propagated wavefields  \citep{LIN97}. At a given frequency $\omega$, we define two integrals:
\begin{equation}
\Psi_{\pm}(\bm{r}, \omega) = \int_{P} \psi(\bm{r}', \omega) H_\pm(\bm{r}, \bm{r}',  \omega)\   \mathrm{d^2}\bm{r}', \label{eq.hographic_image}
\end{equation}
where
\begin{eqnarray}
H_+  &=& - 2 \ii {\rm Re}[k_n] G_0^*,  \qquad  \textrm{ (egression or PB integral)}  \\
H_- &=& \ii k_n G_0  , \qquad\qquad\quad\; \textrm{(ingression)}
\end{eqnarray}
and $G_0$ is the causal Green's function:
\begin{equation}
 - \left[\nabla^2 + k_0^2(\br,\omega)\right]  G_0(\br,\br', \omega)  =   \delta (\br-\br'). \label{eq.greens}
\end{equation}
The wavenumber $k_0$ is computed with Eq.~(\ref{eq.wavenumber}) using the density $\rho_0(r)$ and sound-speed $c_0(r)$ from a 1D reference solar model \citep[extended Model~S,][]{FOU17}, and $k_n$ is the value of $k_0$ at the computational boundary. Equation~(\ref{eq.greens}) is solved by using  the finite-element solver Montjoie \citep{Durufle2006, CHA16, GIZ17}  with a radiation boundary condition applied 500~km above the  solar surface \citep[][equation~7]{GIZ18}.
The first integral, known as the egression, is an estimate of the back-propagated wavefield at the target location $\br$. This definition, which differs from that of \citet{LIN97} by a multiplicative factor (and by the definition of the Green's function), is based on the Porter-Bojarski integral used in the field of  acoustics
\citep[PB integral, see][their equation~14]{GIZ18}. 
The ingression is an estimate of the forward-propagated wavefield at the same location.
Throughout this paper, the domain of integration, $P$, includes all the points from the center of the solar disk to angular distance  $75^\circ$.

Figure~\ref{fig1} shows two example Green's functions in the frequency range  $2.5$--$4.5$~mHz.  
The two-skip Green's function used by the JSOC/Stanford farside imaging pipeline (left panel) is computed under the ray-theory formalism of \citet{LIN97} and the corrections for dispersion described by \citet{LIN00a}. The two-skip Green's function is limited to separation distances in the range  $115^{\circ}<\Delta< 172^{\circ}$  and it is used to image the far side to within a distance of $50^\circ$ from the antipode of disk center \citep[for a graphical representation of the $2\times 2$ skip geometry, see][]{BRA01}. For distances from $50^\circ$ to $90^\circ$ from the antipode (the remaining part of the far side), a $3\times 1$ skip geometry is used in the pipeline. The right panel of Fig.~\ref{fig1} shows the Green's function obtained by solving Eq.~(\ref{eq.greens}) used under the formalism of  \citet{GIZ18}. This particular Green's function contains all spherical harmonics in the range $5\leq \ell \leq 45$. It accounts for all skips, all distances, and finite wavelength effects.  {We refer the reader to Sect.~\ref{sec.ray_fem} for a discussion about the consequences  implied by the choice of Green's function to image active regions on the far side.}  

\subsection{Signal and noise}
\label{sec.sig_and_noise}
Due to the stochastic excitation of solar oscillations, the co-variance of the egression and the ingression, hereafter the holographic image intensity, is used in farside imaging \citep{LIN00b}:
\begin{equation}
    I_\omega (\bm{r}) = \Psi^*_{-}(\bm{r}, \omega) \Psi_+(\bm{r}, \omega).
\end{equation}
In practice, holographic image intensities are averaged over a number of positive frequencies in the range $2.5$\,--\,$4.5$~mHz 
to increase the signal-to-noise ratio:
\begin{equation}
    I(\bm{r})=\frac{1}{N} \sum_{i=1}^N   I_{\omega_i} (\bm{r}),
\end{equation}
where $N$ is the total number of frequencies (see Sect.~\ref{sec.data_reduction}).

In general, the quantity $I$  is complex. We define the phase map as follows:
\begin{equation}
 \Phi (\br) = \arg [   I (\br) /  \langle I_{\rm 0}(\br) \rangle   ] , \qquad {\rm (signal) } 
\end{equation}
where $\langle I_{\rm 0} \rangle$ is a smooth  reference,  chosen to be the mean of the daily $I$ over the quiet-Sun month of  February 2019. 
The phase is computed with the routine \verb!angle! from NumPy and takes values between $-\pi$ and $\pi$.
The level of realization noise is given  by the standard deviation of the phase measured during the quiet-Sun period (using realizations with the same observation duration as for $I$  \citep{GIZ18}:
\begin{equation}
 \sigma(\br)  = \left( {\rm Var} \left[  \arg I_0 (\br) \right] \right)^{1/2} . \qquad {\rm (noise) } 
\end{equation}
The signal-to-noise ratio is then $SNR =|\Phi|/\sigma$.

\begin{figure*}[!htb]
\begin{center}
\includegraphics[width=\linewidth]{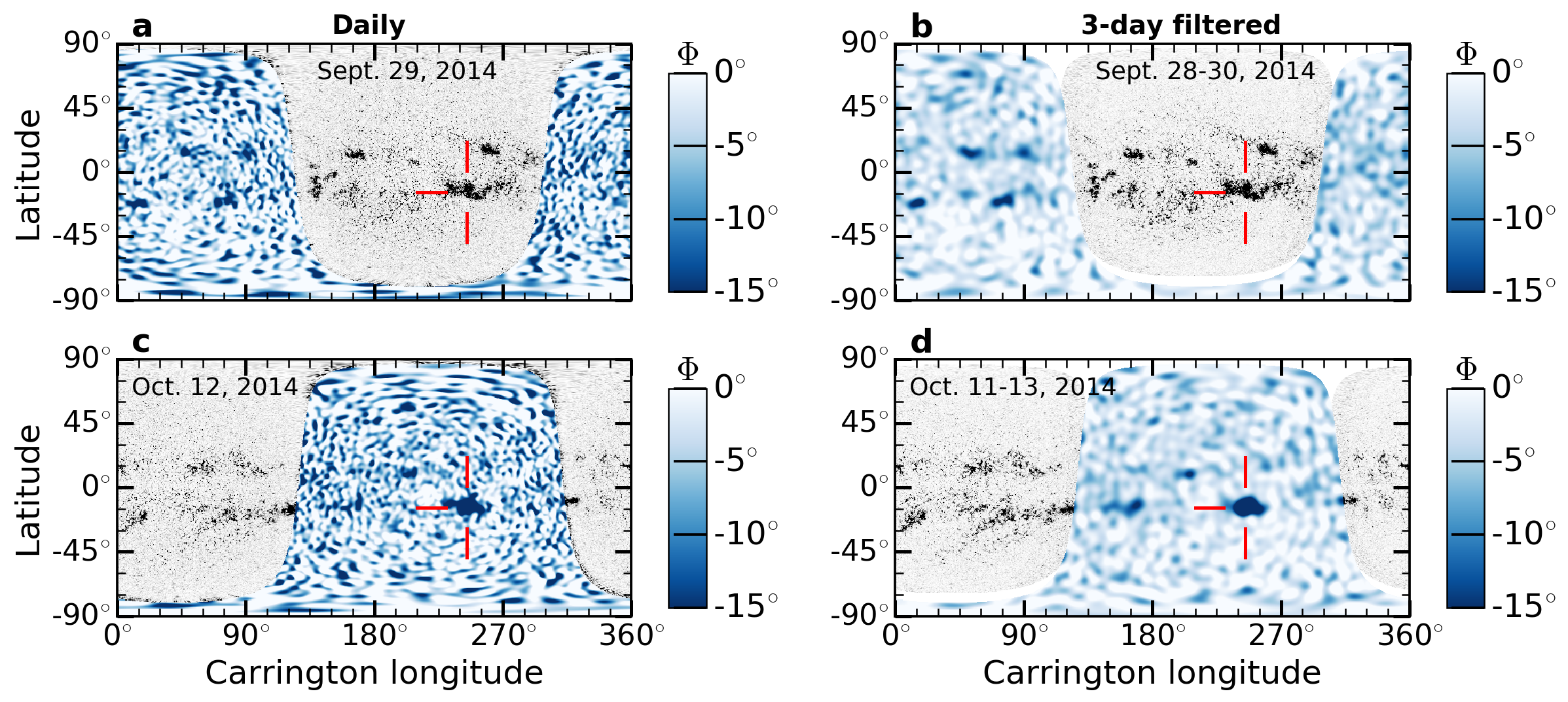} 
\caption{
Composite maps showing the very large active region NOAA~12192. Showed are composite maps, but for different Carrington rotations. STEREO observations are not available for comparison.
{\it Panels (a) and (b)}: 
Seismic map for Sept 29, 2014 (constructed as Fig.~\ref{fig2}a) and 
average over Sept 28--30
(constructed as Fig.~\ref{fig2}b). The longitude is that of Carrington rotation~2155.
{\it Panels (c) and (d)}: 
Same as above, but 13 days later. Active region NOAA~ is very clearly seen on both the near and far sides.
} 
\label{fig3}
\end{center}
\end{figure*}

\begin{figure*}[!htb]
\begin{center}
\includegraphics[width=\linewidth]{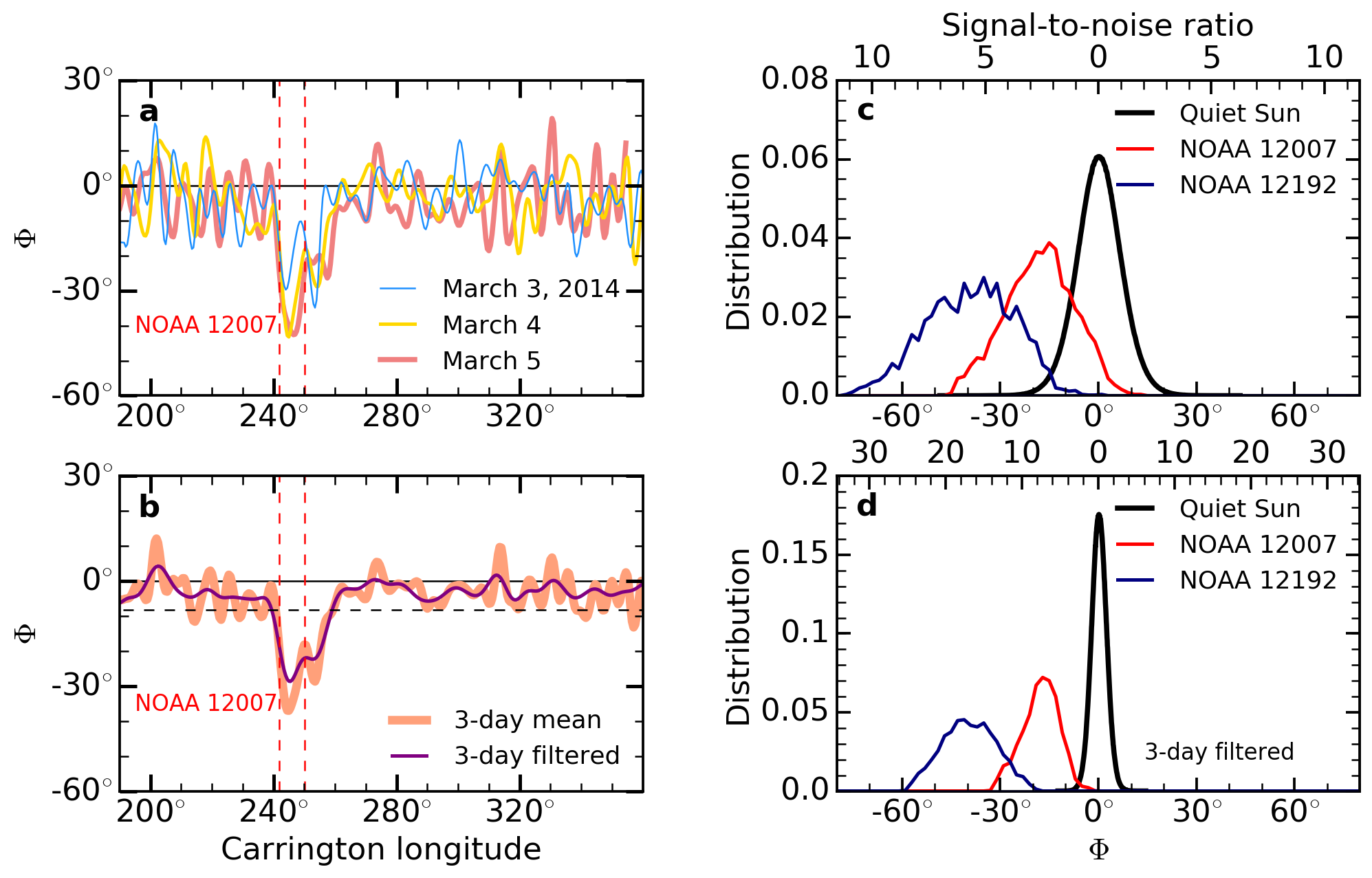} 
\caption{  
Detection of active regions on the far side with helioseismic holography.  
{\it Panel (a)}: Cuts through the phase maps of March 2014 at the latitude of active region NOAA 12007 ($\lambda= 13.5^\circ$, see red horizontal lines in Fig.~\ref{fig2}).
The three curves correspond to  March 3, 4 and 5.
{\it Panel (b)}: The thick orange curve shows the phase when applying holography to three days at once (March 3--5). The purple curve is obtained after applying a spatial low-pass filter ($l<40$). 
{\it Panel (c)}: The red curve shows the normalized distribution of the daily values of $\Phi$ in a disk of diameter $8^\circ$ ($\sim\!100$~Mm) centered on AR NOAA~12007 (see vertical dashed lines in panel a). In total, 14 days data are used to track the region across the far side. The blue curve is the distribution for the very large region NOAA~12192 in October 2014 (see Fig.~\ref{fig3}), constructed in the same way as for NOAA~12007. The black curve is the distribution of daily values in the quiet Sun, from which we measure the standard deviation of the noise, $\sigma=6.9^\circ$. 
{\it Panel (d)}: Same as panel (c), but the distributions are constructed from three-day spatially-filtered maps of $\Phi$, hence the smaller noise level $\sigma=2.3^\circ$. The threshold $th=-3.5 \sigma$ is plotted as a horizontal dashed line in panel (b). We see that NOAA~12007 corresponds to a signal that is much beyond this threshold, with $SNR\sim7$. The very large active region NOAA~12192 is detected with a confidence level of nearly 100\% ($SNR\sim 17$).
}
\label{fig4}
\end{center}
\end{figure*}

\begin{figure*}[!htb]
\begin{center}
\includegraphics[width=\linewidth]{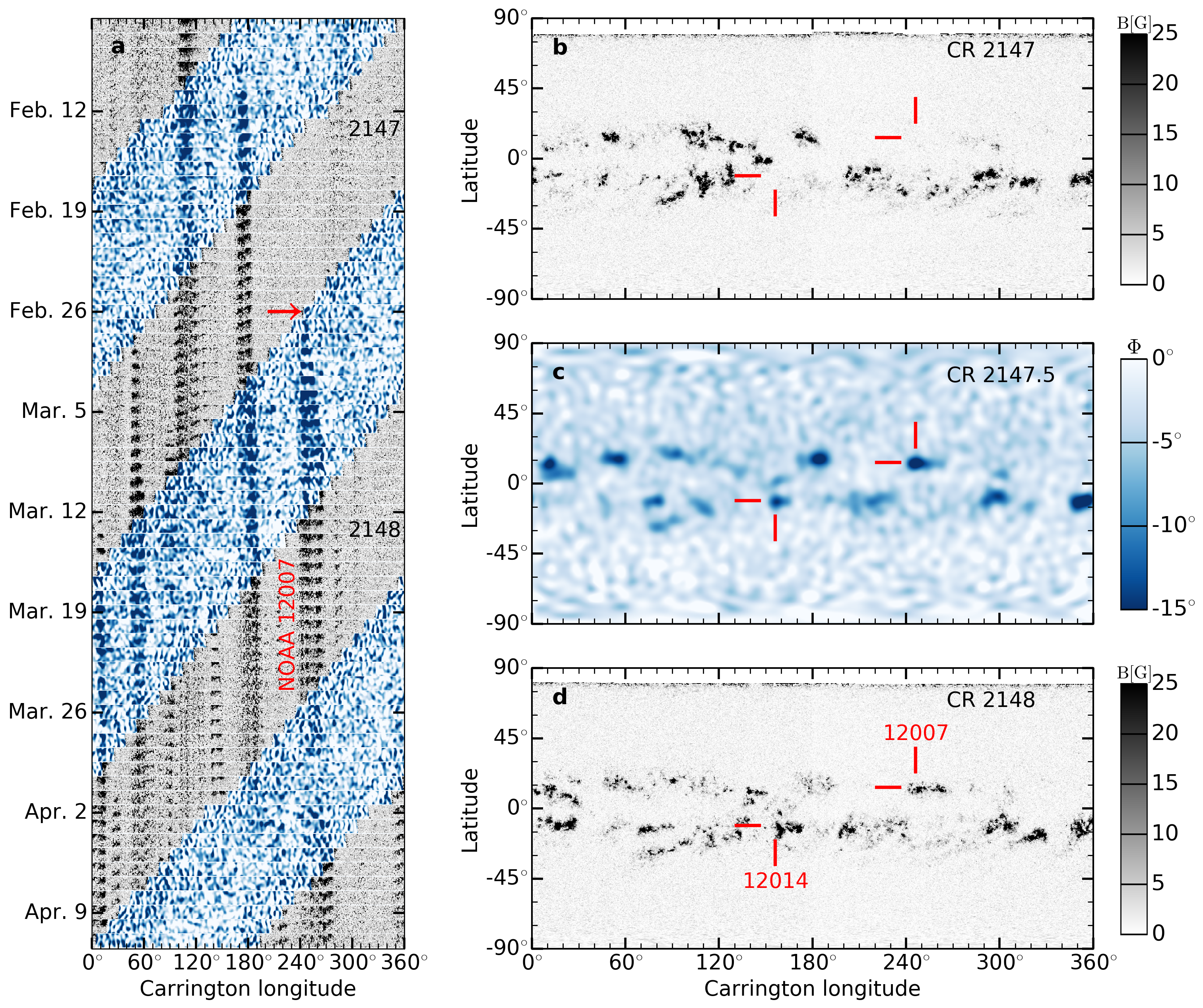} 
\caption{
Emergence of active regions on the far side.
{\it Panel (a)}: Longitude--time diagram extracted from the daily composite maps of Fig.~\ref{fig2}a. Narrow bands in latitude ($10^\circ < \lambda < 20^\circ$) are stacked together to show the evolution of individual active regions. The blue shades correspond to the far side (daily seismic phase), and the gray shades to the near side (magnetograms, from CR~2147 to CR~2148).
The red arrow indicates the time when active region NOAA~12007 emerges, near the limb on the far side. 
{\it Panels (b)-(d)}: Carrington synoptic maps constructed by averaging the magnetograms on the near side (CR~2147 and~2148) and the seismic phase on the far side (CR 2147.5). At any spatial location, the data are averaged over $~14$~days in these Carrington maps. The red lines point to two particular active regions, NOAA~12007 and 12014, which are not visible in CR~2147 but visible in CR~2147.5 and~2148.} \label{fig5}
\end{center}
\end{figure*}

\section{Data reduction}
\label{sec.data_reduction}

Helioseismology instruments do not provide us with direct access to the quantity $\psi$ defined in Eq.~(\ref{eq.psi}). Instead, we may consider rough approximations of $\psi$, using either intensity images or Dopplergrams. Here we used  full-disk Dopplergrams from  SDO/HMI, each with $4096 \times 4096$ pixels and a spatial sampling of $0.5$~arcsec and  taken once every 45~s \citep{SCH12}.  Overlapping segments of 31~hr were processed with a cadence  of  24 hr. 
The data set that we analysed covers the period from May 1, 2010 to December 31, 2019.
We first rebined the individual Dopplergrams over $4\times4$ pixels, and subtract a smooth fit.  
Each Dopplergram was divided by the sine of the angle between the local normal to the Sun's surface and the line of sight, in order to correct the amplitude variation due to projection.  We then  Postel projected each Dopplergram onto a grid with pixels separated by $0.002$~rad (using bicubic interpolation). We obtained sets of 31-hr data cubes ($2480$  time steps) that track the Sun in the Carrington frame. The Dopplergrams were further binned by a factor of 8 along each spatial dimensions, and were detrended at each pixel to remove the mean and the slope in time for each data cube. Values five sigmas away from the mean (zero) were set to zero.  The data cubes were transformed  with respect to time using a fast Fourier transform.  {To circumvent the assumption of periodicity}, the data sets were zero padded in time  ($4096$ time steps), implying  a frequency resolution of $5.425$~$\mu$Hz. A bandpass filter was applied to select frequencies in the range from $2.5$ to $4.5$~mHz. 

\section{Results}
\subsection{Example farside maps}
\label{sec.results}
Figure~\ref{fig2} shows example composite maps of the solar surface in the Carrington  frame for rotation CR~$2147$ in March 2014. In all four panels,  we show the unsigned magnetic field  $B_\textrm{los}$  from SDO/HMI on the near side. Fig.~\ref{fig2}a shows  the seismic phase $\Phi$ on the far side from one data segment of 31~hr centered on 4 March 2014, using the improved wave propagators. 
The phase map captures the largest active regions that are also seen in the 304 \si{\angstrom} STERERO/EUVI image on the same day on the far side (Fig.~\ref{fig2}c), albeit with a rather low signal-to-noise ratio.
Fig.~\ref{fig2}b is similar to Fig.~\ref{fig2}a, but the seismic phase on the far side is  averaged over about three days. More precisely, we average $I$ and $I_0$ over three overlapping 31-hr time segments centered respectively on March 3, 4, and 5, 2014,  we compute $\Phi$, and we filter out  the small scales (spherical harmonic degrees $l > 40$). This 79~hr averaged map (Fig.~\ref{fig2}b) reveals the presence of relatively small active regions. By comparison, the 79~hr averaged map from the JSOC/Stanford  pipeline (Fig.~\ref{fig2}d) appears to be  more  blurred. 

Figure~\ref{fig3} shows the detection of a very large active region NOAA~12192 on both the near and far sides of the Sun. Fig.~\ref{fig3}a and ~\ref{fig3}b are constructed the same way as in Figs.~\ref{fig2}a and \ref{fig2}b, but for a daily maps on Sept. 29, 2014 and three-day average over Sept. 28--30. The large active region NOAA~12192 is clearly visible on the near side (red segments), whereas its signature is evidently captured by the phase map 13 days later on the far side (Fig.~\ref{fig3}c and ~\ref{fig3}d).

\subsection{Active-region detection confidence levels}
Figure~\ref{fig4} shows further details on how to distinguish active regions from the noisy background. Figs.~\ref{fig4}a and \ref{fig4}b  show slices through the active region NOAA~12007 shown in the maps of $\Phi$ along the latitude marked by the red horizontal segments shown in Figs.~\ref{fig2}a and \ref{fig2}b. The daily slices are plotted in Fig.~\ref{fig4}a, and the averages over three days are shown in Fig.~\ref{fig4}b. 
We clearly see the signature of this medium-size active region in the phase $\Phi$, however there are fluctuations due to p-mode realization noise. The signal-to-noise ratio improves when averaging over three days and filtering out the small spatial scales (see Fig.~\ref{fig4}b).

To quantify the detection confidence level, we show in Figs.~\ref{fig4}c and \ref{fig4}d 
the distributions of the values of $\Phi$  within two active regions, NOAA~12007  and NOAA~12192 (a huge region), and we compare these distributions to the quiet Sun distributions (pure noise).
The detection confidence level is high when there is little overlap between the distribution of values in an active-region and the  distribution corresponding to noise.
For one day of data (Fig.~\ref{fig4}c), we find that NOAA~12192 is detected with a confidence level of nearly  $100 \%$, while NOAA~12007 is detected with a signal-to-noise ratio of about $2.5$ (defining the signal to noise ratio as the ratio between the mean of the active region distribution and the standard deviation of the noise, see scale on the top axis of Fig.~\ref{fig4}c). 
For three days of data (Fig.~\ref{fig4}d), both active regions are detected with a confidence level of nearly  $100 \%$ (signal-to-noise ratios of 7 and 17). Thus time averaging and spatial filtering is an efficient mean to improve the detection level of medium- and small-size active regions that do not evolve too fast.

\subsection{Daily evolution of active regions on the far side}

To follow the daily  evolution of individual active regions, from their emergence until they fade away, we cut slices at constant latitudes through the maps in Fig.~\ref{fig2}a. An example longitude-time map is shown in Fig.~\ref{fig5}a, where daily slices with latitudes from $10^\circ$ to $20^\circ$ (in the North) are shown from the beginning of CR 2147 to the end of CR 2148. This map shows the evolution of individual active regions in this latitude range. In particular, we see that the active region NOAA 12007 emerges on the far side near the solar limb on 26 February 2014 (as indicated by the red arrow), grows in size  into a bipolar active region, and starts to decay after it appears on the near side.\footnote{We refer the reader to the time-distance helioseismology results by \citet[][their figure 9]{ZHA19}, where active region NOAA~12007 is detected in 2-day averages around 6 March 2014.} 
We also compare these seismic maps with STEREO/EUVI observations of the far side, and find that they show a very similar evolution of the active regions  (see Figs.~\ref{figa2} in the north and \ref{figa3} in the south).
This demonstrates that seismic maps can not only be used to detect newly emerged  regions on the far side, but can also be used to characterize their evolution day after day.

\subsection{Emergence of active regions on the far side}
\label{sec.4.4}
The right panels in Fig.~\ref{fig5} show three Carrington synoptic maps of the solar surface, which illustrate how to identify active regions that emerge on the far side. The top and bottom maps (Figs.~\ref{fig5}b and d) are for Carrington rotations CR~2147 and CR~2148 respectively; they were constructed using  line-of-sight magnetograms of the near side. The synoptic map in the middle (Fig.~\ref{fig5}c) was constructed using the seismic phase on the far side and is denoted CR~2147.5. 
In each map the value at a spatial location corresponds to a time average over approximately two weeks \citep[see also, e.g.,][]{GON07}.  These three consecutive maps provide a picture of solar activity at times that are separated by approximately 13.5 days from one another (at fixed Carrington longitude).

Due to the long time averaging, active regions on the far side produce a sharp signature in the synoptic maps and are quite easy to distinguish from background noise.
The spatial resolution is about twice the diffraction limit, or about 100~Mm. 
In the synoptic maps, we see a number of active regions at similar spatial locations on all three consecutive maps (Figs.~\ref{fig5}b--d), which indicates their long lifetimes.
Some other regions are not present on all three maps.
For example, the two active regions highlighted by the red line segments in the figure are seen on the maps for CR 2147.5 and CR~2148, but they are not seen on the previous map CR~2147 (Fig.~\ref{fig5}b). 
The active region in the North emerged on the far side (Fig.\ref{fig5}~a) and can be identified later on the near side as active region NOAA 12007.
The active region in the South emerges first near the limb as NOAA~11995, grows across the far side, and reappears on the near side as NOAA~12004 (Fig.~\ref{figa3}).

Synoptic maps from other Carrington rotations are shown in Fig.~\ref{figa4}, during both solar maximum and solar minimum. It is reassuring that the farside maps show no false positive detections during quiet times. This means in particular that farside maps may provide important information for all-clear space weather predictions.

\begin{figure}[!htb]
\begin{center}
\includegraphics[width=\linewidth]{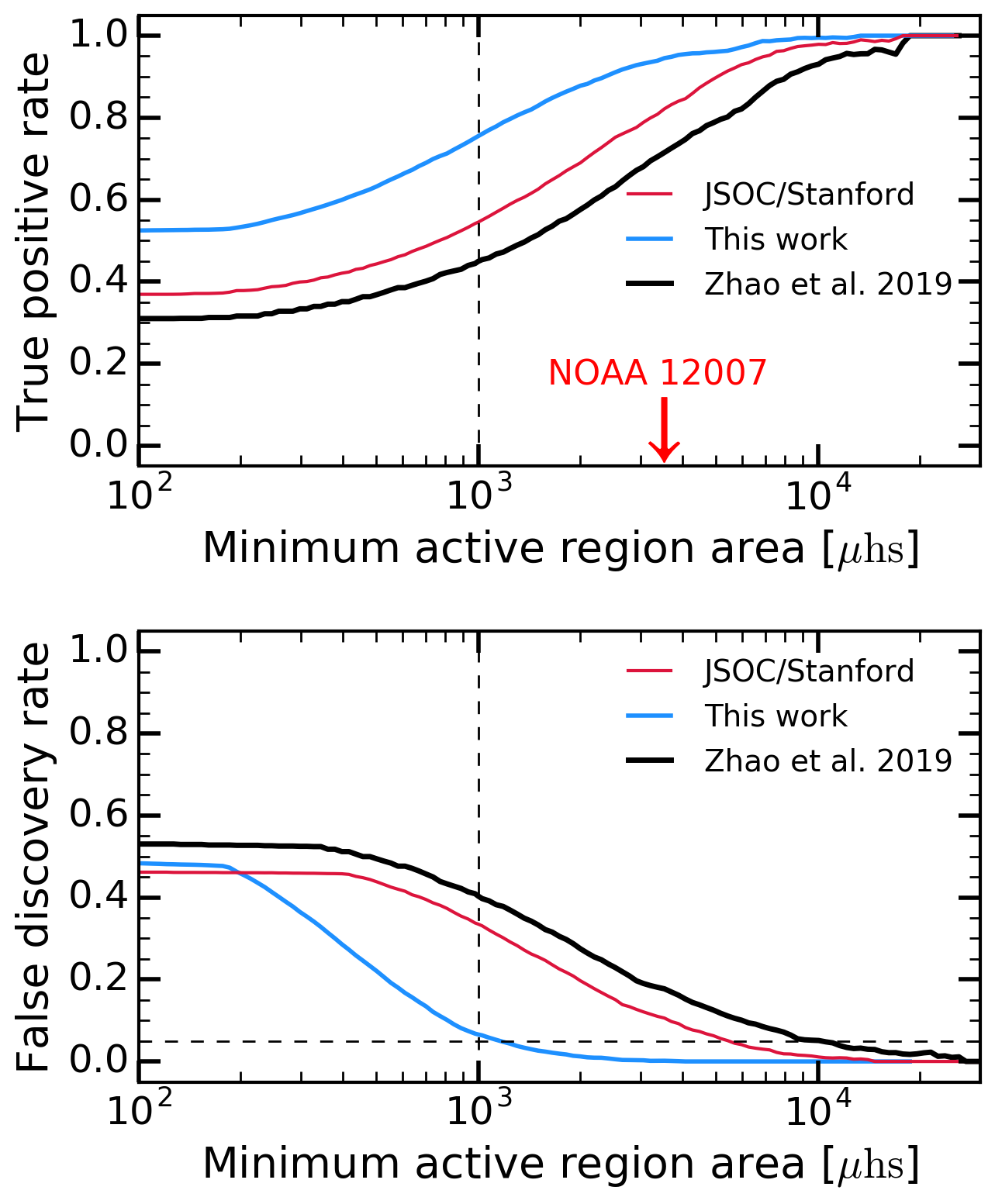}  
\end{center}
\caption{Active region detection statistics on the far side. The seismic maps from  this work (3-day averages), the JSOC/Stanford seismic holography
‘strong region maps' (5-day averages), and the maps from time-distance helioseismology  \citep[][4-day averages]{ZHA19}  are compared with STEREO maps (4-day averages).
{\it Top panel}: True positive detection rates for the three seismic imaging methods as a function of minimum active region area  $S_{\rm min}$. The area of NOAA 12007 is $S= 3.4\times 10^3$~$\mu\rm{hs}$ (red arrow and Fig.~\ref{fig2}b). 
{\it Bottom panel}: False discovery rates. The horizontal dashed line is drawn at $5\%$. 
Our improved method has a significantly smaller false discovery rate than the other methods, in the  $S_{\rm min}$ range of interest.
}\label{fig6}
\end{figure}

\subsection{Comparisons to other seismic methods}
\label{sec.4.5}
STEREO observations of active regions on the far side have  frequently been used as a reference to evaluate the validity of seismic images. We use STEREO/EUVI 304~\si{\angstrom} maps of chromospheric emission remapped in a Carrington frame from \url{http://sd-www.jhuapl.edu/secchi/jpl/euvisdo_maps_carrington/304fits} (averaged over 4 days) and we apply the thresholding method described by \citet{LIE17} to identify individual active regions on the far side.
Example STEREO maps for October 2013 are shown in Fig.~\ref{figa5}a and the detected active regions as black regions in Fig.~\ref{figa6}a. The rectangular boxes around the detected active regions have areas proportional to the active region size.

In order to extract active regions from our seismic maps (3 day averages with Gaussian smoothing, Fig.~\ref{figa5}b), we apply a threshold to the phase $\Phi$ of $th = -3.5$ times the standard deviation $\sigma = 0.045\ \textrm{rad} = 2.6^\circ$  measured from quiet-Sun maps. Following \citet{LIN17}, we remove all detected regions that have an area $S$ such that the integrated phase  $|\int_S \Phi dS|$ is less than $30$~rad per millionth of an hemisphere ($1\ \mu\textrm{hs} = 10^{-6} \times 2\pi R_\odot^2$). Typically, this means that regions that are smaller than about $5$ square degrees on the solar surface are removed. The results are shown in Fig.~\ref{figa6}b. The same algorithm (with $th = -3 \sigma$, $\sigma = 0.023\ \textrm{rad} = 1.3^\circ$) is used to identify active regions from the JSOC/Stanford seismic holography `strong region maps' (5 day averages with Gaussian smoothing, Fig.~\ref{figa5}c) available at \url{http://jsoc.stanford.edu/data/farside/Phase_Maps_5Day_Cum}. The corresponding active regions are shown in Fig.~\ref{figa6}c. 
For comparison, we also use the farside maps obtained with time-distance helioseismology by \citet{ZHA19}. These maps are available at  \url{http://jsoc.stanford.edu/data/timed} (4 day averages, \verb$hmi.td_fsi_12h$ in the JSOC DRMS, see Fig.~\ref{figa5}d). 
Using the threshold in $\Phi$ of $-0.06$~rad  proposed by \citet{ZHA19} and removing the regions that contain no more than 10 pixels, we obtain the active regions shown in Fig.~\ref{figa6}d.

Assuming that the STEREO active regions  represent the ground truth, we wish to assess the active region detection rates for the three types of seismic maps mentioned above. 
We consider an active region detection to be true if the active region detected with STEREO is also detected in a seismic map, i.e. if the two active region areas overlap. A detection is said to be false if an active region seen in a seismic map  is not seen in the  STEREO map, i.e. if there is no overlap between active region areas. We refer the reader to, e.g., \citet{TIN10} and \citet{YOA95} for the definitions of the true positive and false discovery rates.

Figure~\ref{fig6} shows the rates of true positives and false discovery rate for the different seismic  methods. The detection rates are applied to sets of active regions with areas above a certain value $S_{\rm min}$. In total, two and half years of data are used from 1 January 2012 to 30 June 2014, which corresponds to the period during which the two STEREO satellites observed a large fraction of the far side. 
We find that the JSOC/Stanford holography maps and the  time-distance maps return very similar detection rates. The holography maps from this paper lead to a significantly higher true positive detection rate and a significantly smaller false positive detection rate than the other two methods. For a minimum active region area  $S_{\rm min}=1000$~$\mu\mathrm{hs}$, the true positive rate reaches $\sim\!75~\%$ and the false discovery rate drops to $\sim\!7\%$.
This means  that a proper  calculation of the Green's function matters in this problem.

\begin{figure}[!t]
\begin{center}
\includegraphics[width=0.9\linewidth]{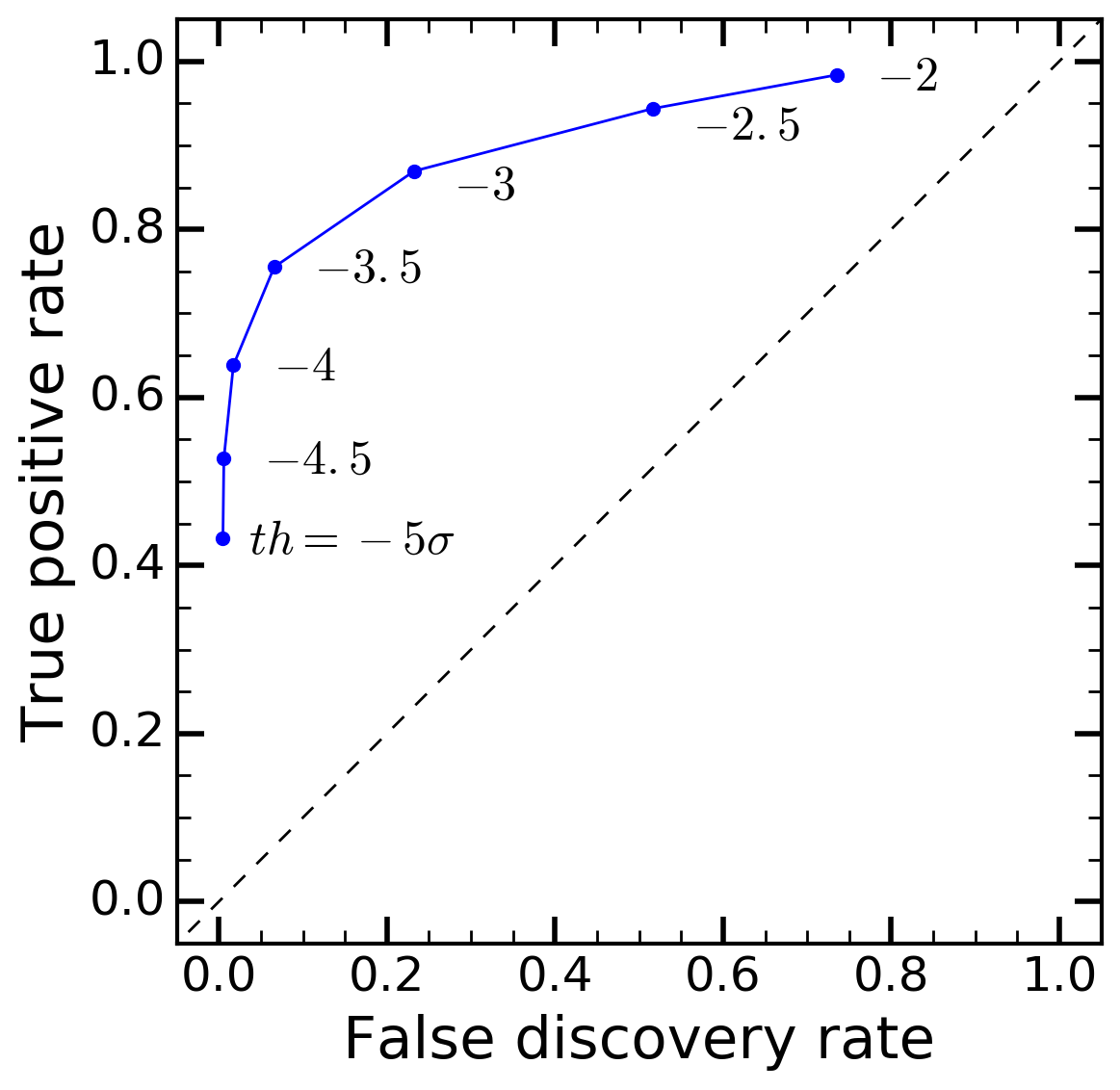}  
\end{center}
\caption{ROC analysis: true positive  rate versus false discovery rate for different thresholds in  $\Phi$ ($th$). Our improved farside maps (3-day averages) are used and the minimum active-region area is fixed at $S_{\rm min}=1000$~$\mu\mathrm{hs}$. Random detections would  fall on the black dashed line.  For all thresholds $th$, the outcome is clearly above this diagonal. 
The optimal threshold, $th = -3.5\sigma$, corresponds to the point that is farthest from the diagonal.
It is reported as a horizontal dashed line in Fig.~\ref{fig4}b. } \label{fig7}
\end{figure} 

Our choice of threshold, $th=-3.5\sigma$ for 3-day averages, can be justified {\it a posteriori} using a Receiver Operating Characteristic (ROC) analysis \citep[see, e.g.,][]{FLA10}.
Figure~\ref{fig7} shows the true positive rate versus the false discovery rate for  several values of the threshold $th$ from $-5\sigma$ to $-2\sigma$, while fixing the minimum active region area to $S_{\rm min} = 1000$~$\mu\mathrm{hs}$. 
We see that all values of $th$ give a significantly better outcome than a purely random detection method (black dashed line with slope 1). 
The optimal threshold lies between $th=-3.5\sigma$ and $th=-3 \sigma$, where the distance to the diagonal (corresponding to a purely random detection method) is the largest in the ROC diagram.

\begin{figure*}[!htb]
\begin{center}
\includegraphics[width=\linewidth]{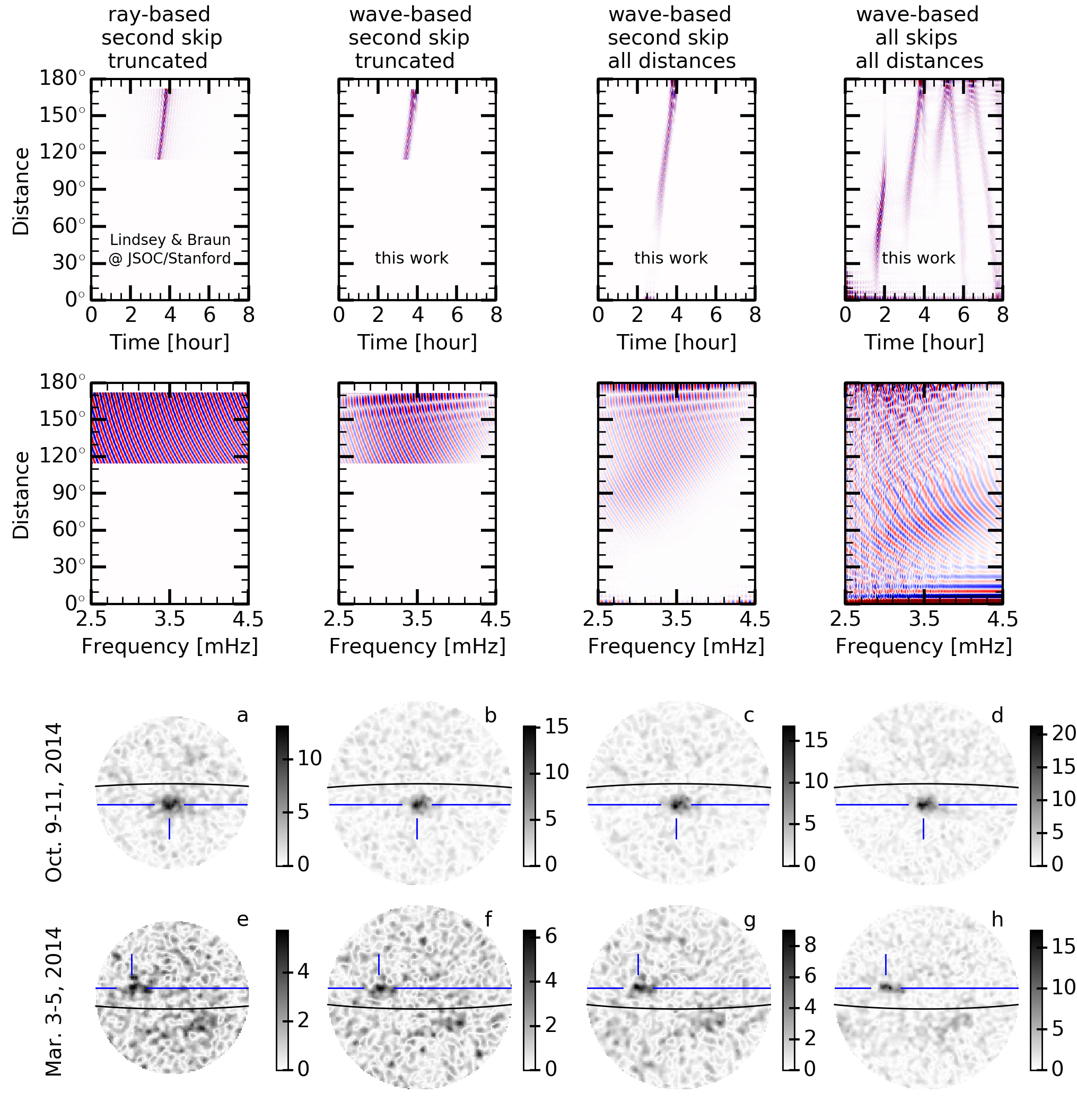}
\caption{
 {
Green's functions and associated signal-to-noise ratios of  phase maps of the far side. 
{\it Top row}: Green's functions in the time--distance domain.
{\it  Second row}: Imaginary part of  Green's functions in the frequency-distance domain. 
{\it Third row}: Signal-to-noise ratios for active region NOAA~12192 averaged over  9--11~October 2014.
{\it  Last row}: Signal-to-noise ratios for active region NOAA~12007 averaged over 3--5~March 2014.
{\it Panels (a) and (e)}: Maps using the ray-based Green's function from the JSOC/Stanford pipeline, for focus points within 50$^\circ$ of the farside disc center. 
{\it Panels (b) and (f)}: Maps using the wave-based Green's function restricted to the second-skip and separation distances 115$^{\circ}-$172$^{\circ}$. The map covers 60$^\circ$ within the farside disc center. 
{\it Panels (c) and (g)}: Same as panels b and f, but without truncating the Green's function in separation distance. 
{\it Panels (d) and (h)}: Same as panels c and g, but using all branches of the wave-based Green's function, i.e. the approach used in the present paper.  
In panels a--h, the black solid lines mark the Equator. The blue lines indicate locations of active regions NOAA~12192 (third row) and NOAA~12007 (last row).} 
} 
\label{fig_snr_postel}
\end{center}
\end{figure*}

\subsection{ {Where does the improvement come from? }}\label{sec.ray_fem}
 {
We have shown that the Green's function that solves Eq.~(\ref{eq.greens}) leads to a significant  improvement in  helioseismic farside imaging. 
This Green's function  differs from the simplified Green's function used  by  \citet{LIN00a} in several ways (see Fig.~\ref{fig1}): (i) all branches (all skips) are kept, 
(ii) all separation distances are kept, and (iii) the details of the solution in each particular branch differ (at least for the second skip).
To find out about the contribution of these differences  to the improvement, we apply several truncations to our Green's function and compare the results to those of the JSOC/Stanford pipeline.

First, we keep only the second skip part of the Green's function (for the branch with the shortest path) and we truncate the separation distances as in the JSOC/Stanford pipeline ($115^\circ<\Delta<172^\circ$).  The resulting signal-to-noise ratio of active regions in the helioseismic maps is shown in the second column of Figure~\ref{fig_snr_postel} for two particular active regions. The improvement is only $\sim 10$\%, implying that the wave-based Green's function is not significantly different from the ray-based function \citep[once corrected using the `local control  correlation', see][]{LIN00a}.

In a second step, we study the consequences of using all separation distances while keeping the second skip only. The corresponding signal-to-noise ratios are shown in the third column of Fig.~\ref{fig_snr_postel}. We see that filling out the hole in the observation pupil leads to better focusing on the far side. The signal-to-noise ratio increases by $\sim 10$\% when including all available data in the analysis (see the changes from Figs.~\ref{fig_snr_postel}b and~f to  Figs.~\ref{fig_snr_postel}c and~g).

In a third step, we look at the relative improvement brought by letting through the other branches of the Green's function (Fig.~\ref{fig_snr_postel}, top right panel). The corresponding changes are seen in the transition from Figs.~\ref{fig_snr_postel}c and~g to  Figs.~\ref{fig_snr_postel}d and~h).
The additional branches of the Green's function clearly lead to the most significant improvement in the signal-to-noise ratio, at a level of $50$\% to $100$\% depending on the active region.

A concise view of the three successive improvements is presented in Figure~\ref{fig_snr_postel_slice}.  Cuts through the active regions in the seismic maps (phases) confirm that the improvement is mostly due to the use of all the branches in the Green’s function and the use of a full observation pupil (all the available data are used as input).
}

\section{Conclusion}
\label{sec.outlook}

We demonstrated that the spatial resolution and the signal-to-noise ratio of helioseismic holography maps of the Sun's far side improve when accurate propagators are used in the analysis. Here we used the finite-wavelength Green's function computed in the frequency domain by \citet{GIZ17} using the transparent boundary conditions determined by \citet{FOU17} and \citet{Barucq2018}. In particular, we find that a spatial resolution of order 100~Mm can be reached, i.e. twice the size of a sunspot.

When it comes to the detection of active regions, we showed in Sect.~\ref{sec.4.5} that seismic maps averaged over three days provide a detection rate as high as $\sim\!75\%$ and a false discovery rate as small as $\sim\!7\%$ for a minimum active region area  $S_{\rm min} = 1000$~$\mu\mathrm{hs}$.
The outcome depends sensitively on the threshold $th$ used to identify active regions in the phase maps, as illustrated by the ROC diagram (Fig.~\ref{fig7}).
Given that the false discovery rate is very small -- which can also be verified during quiet-Sun periods --  the farside maps should  be most helpful to forecast all-clear space weather  conditions \cite[see, e.g.,][]{HILL18}. 

We also showed that seismic maps can track the emergence and evolution of individual active regions on a daily time scale as they move across the far side. 
This conclusion was reached using STEREO chromospheric images as a reference.
The upcoming photospheric observations of the far side by the Polarimetric and Helioseismic Imager on Solar Orbiter \citep{SOL20} should provide a definitive confirmation of the accuracy of farside helioseismic holography.

\begin{acknowledgements}

We are very grateful to Charles Lindsey 
(NWRA/CoRA) for providing access to his helioseismic holography code and very helpful discussions. We thank Damien Fournier (MPS) for his help with the  computation of the Green's functions. We also thank an anonymous referee for constructive comments.
LG and DY acknowledge funding from the ERC Synergy Grant WHOLE SUN (\#810218) and the DFG Collaborative Research Center SFB 1456 (project C04). LG acknowledges NYUAD Institute Grant G1502.
{HB acknowledges funding from the 2021-0048: Geothermica SEE4GEO of European project  and the associated team program ANTS of Inria.} 
The HMI data are courtesy of SDO (NASA) and the HMI consortium. The SDO/HMI holographic farside images are computed and maintained by JSOC/Stanford at \href{http://jsoc.stanford.edu/data/farside/}{jsoc.stanford.edu/data/farside/}. The SDO/HMI time-distance farside images are  made available at \href{http://jsoc.stanford.edu/data/timed/}{jsoc.stanford.edu/data/timed/}.
The EUVI maps are generated by the SECCHI team and maintained at JHUAPL, in collaboration with NRL and JPL. 
\end{acknowledgements}

\bibliographystyle{aa}
\bibliography{biblio.bib}

\appendix

\section{Additional figures }

\begin{figure*}[!htb]
\begin{center}
\includegraphics[width=\linewidth]{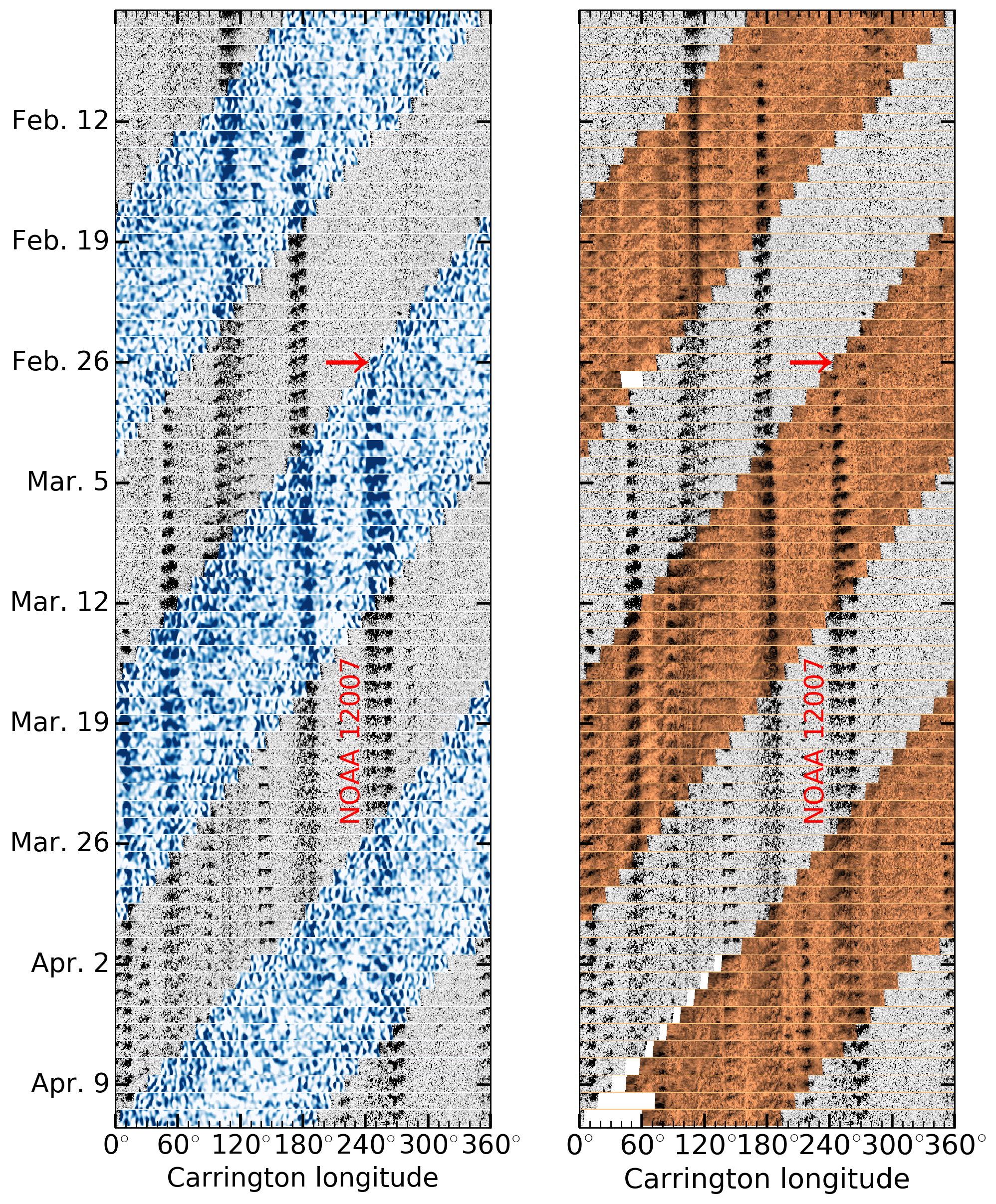}
\caption{
Longitude–time diagrams extracted from the daily composite maps for latitudes bands in the North ($15^\circ \pm 5^\circ$).
{\it Left panel}: 
The farside maps are obtained from seismology and shown in blue  (same as Fig.~\ref{fig2}a). 
{\it Right panel}: The farside maps are obtained from STEREO/EUVI and  shown in orange. The red arrow marks the time when active region NOAA~12007 emerges.} \label{figa2}
\end{center}
\end{figure*}
\begin{figure*}
\begin{center}
\includegraphics[width=\linewidth]{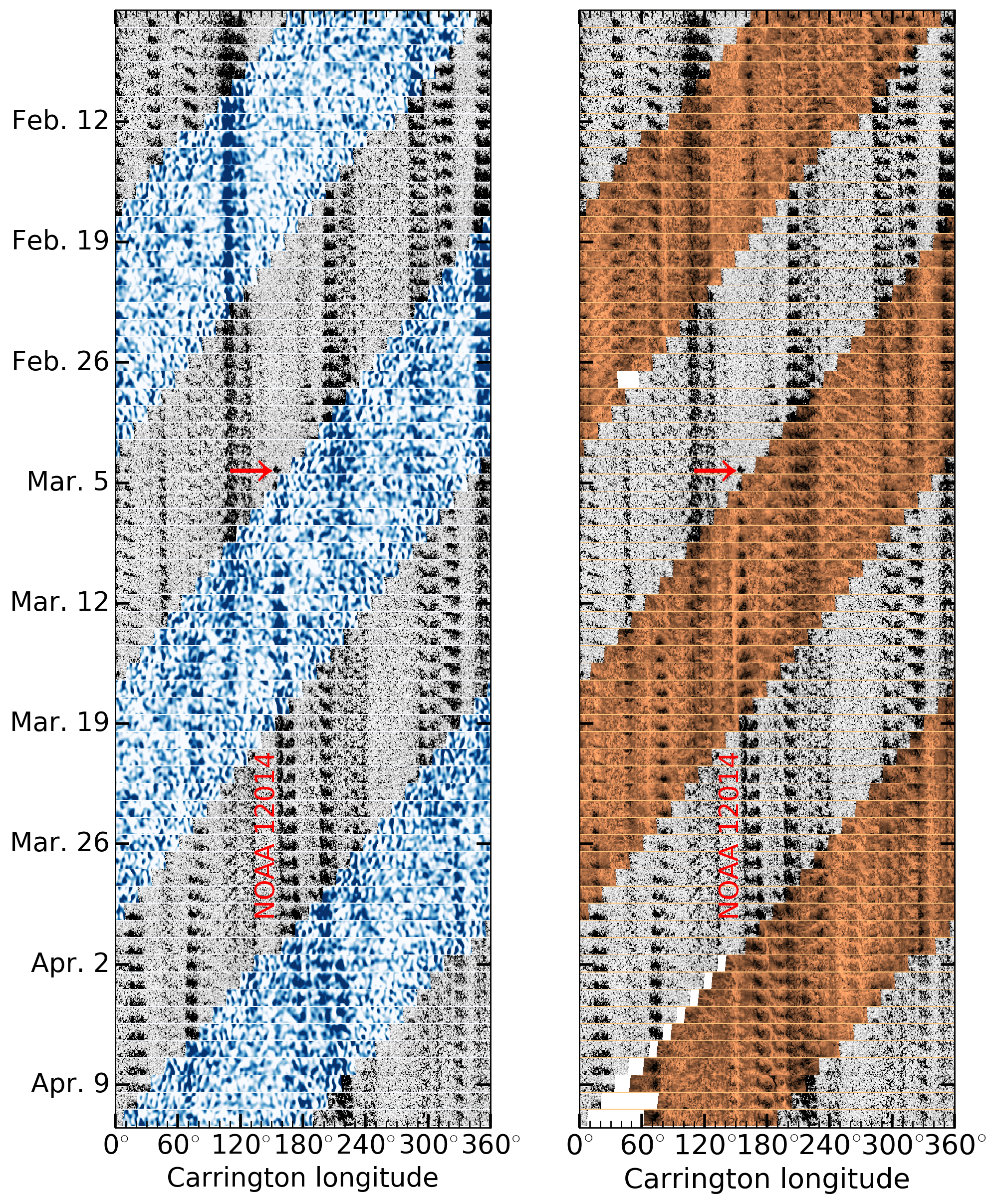}
\caption{Same as Fig.~\ref{figa2}, but for latitudes bands in the South ($-15^\circ \pm 5^\circ$). The red arrow marks the time when active region NOAA~11995 emerges (called 12014 one rotation later).  
} \label{figa3}
\end{center}
\end{figure*}

\begin{figure*}
\begin{center}
\begin{tabular}{cc}
\includegraphics[width=0.45\linewidth]{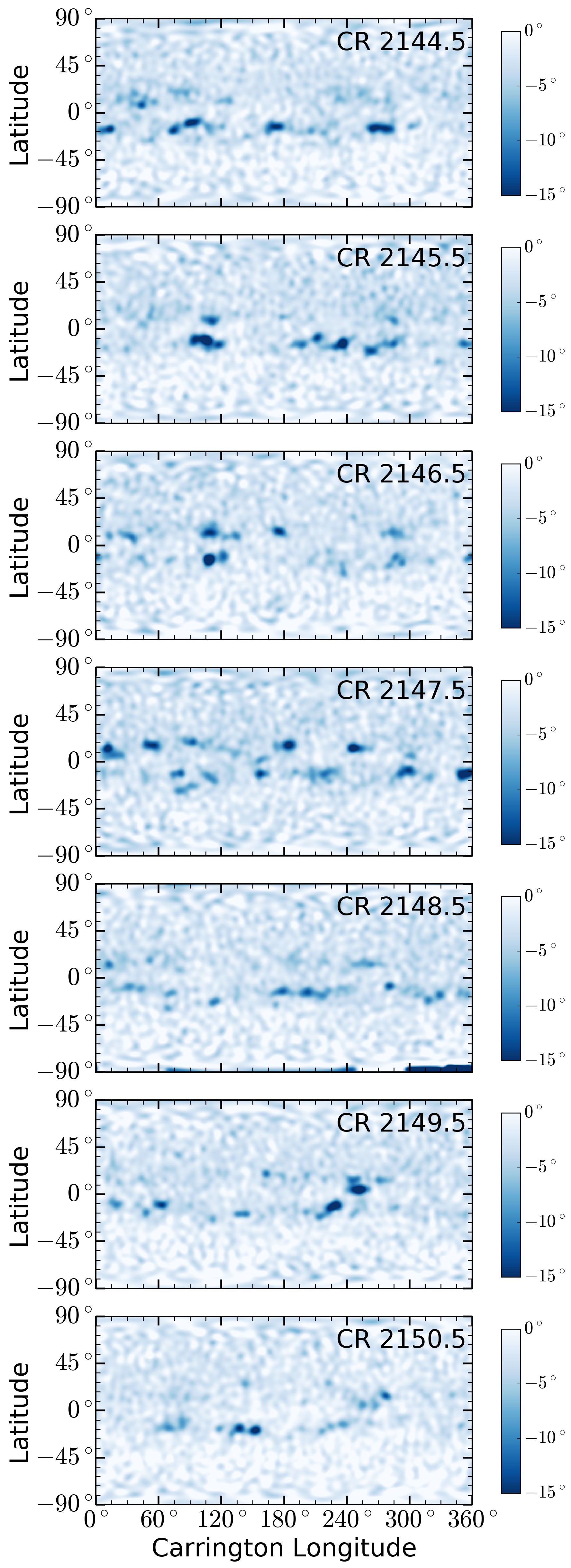} & \includegraphics[width=0.45\linewidth]{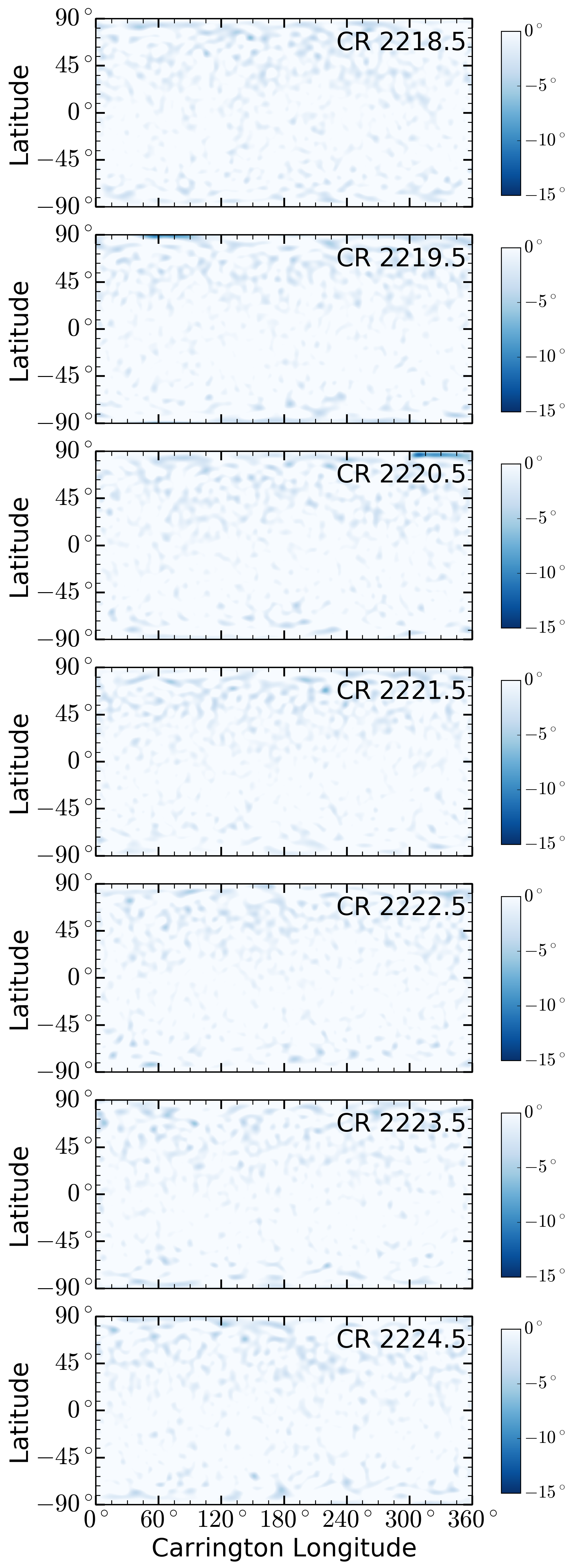} 
\end{tabular}
\caption{Seismic synoptic maps of the far side constructed as in Fig.\ref{fig4}c. {\it Left panels}:  Carrington rotations 2144.5--2150.5 during the maximum phase of  Cycle 24 (2013--2014).
{\it Right panels}: Carrington rotations 2218.5--2224.5 during solar minimum in 2019.
    The same color scale is used for all maps.
} \label{figa4}
\end{center}
\end{figure*}

\begin{figure*}[!htb]
\begin{center}
\includegraphics[width=\linewidth]{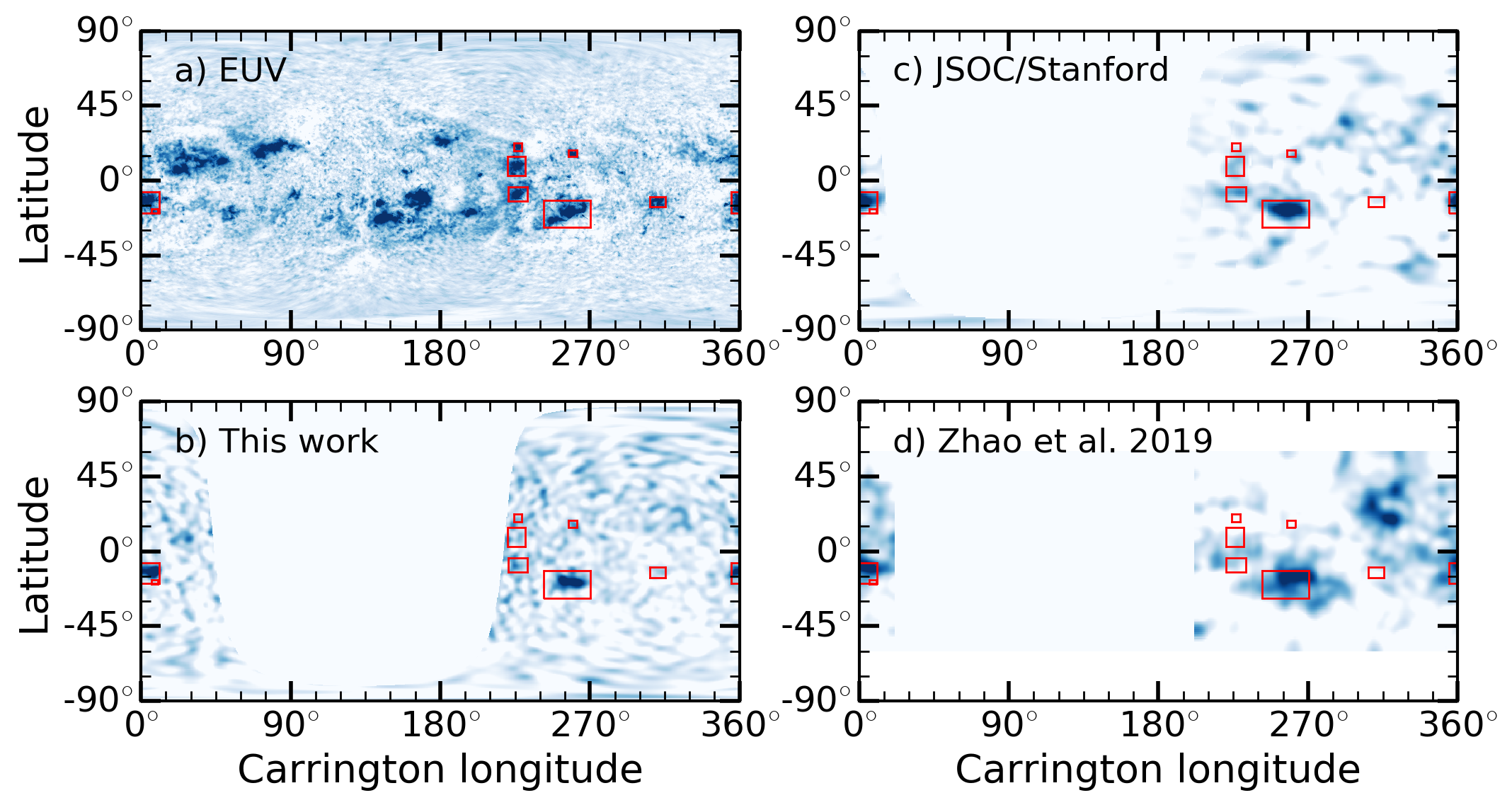}  
\end{center}
\caption{Detections of farside active regions  on 15 October 2013 (12:00 TAI). {\it Panel a}: Composite EUV maps at 304~\si{\angstrom}   from \citet{LIE17} (4-day averages). The detected active regions are enclosed in the red boxes. {\it Panel b}: Seismic farside images from this work (3-day averages). For reference, the red boxes are copied from panel a. {\it Panel c}: Same as panel b but for holographic farside maps from JSOC/Stanford (5-day averages). {\it Panel d}: Same as panel b but for  time-distance farside maps from \citet{ZHA19} (4-day averages).  
} 
\label{figa5}
\end{figure*}

\begin{figure*}[!htb]
\begin{center}
\includegraphics[width=\linewidth]{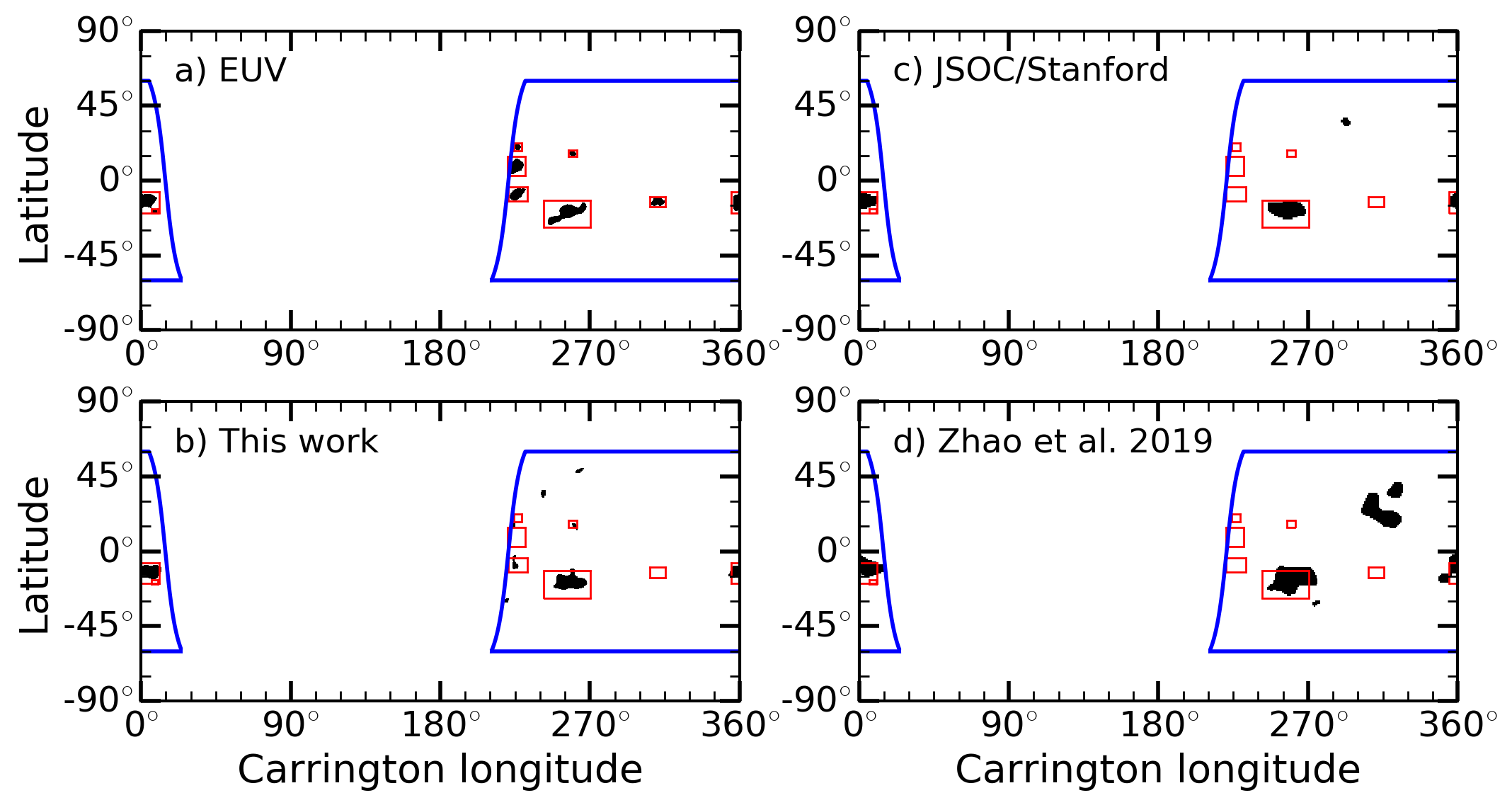}
\end{center}
\caption{Detected active regions in each of the four panels of Fig.\ref{figa5} (black regions). For reference, the red boxes are copied from Fig.~\ref{figa5}a (EUV map).  Latitudes above $60^\circ$ are ignored (cf. blue contours). Such data are used to generate the true positive detection rates and the false discovery rates discussed in Sect.~\ref{sec.4.5}. 
} 
\label{figa6}
\end{figure*}

\begin{figure}[!htb]
\begin{center}
\includegraphics[width=\linewidth]{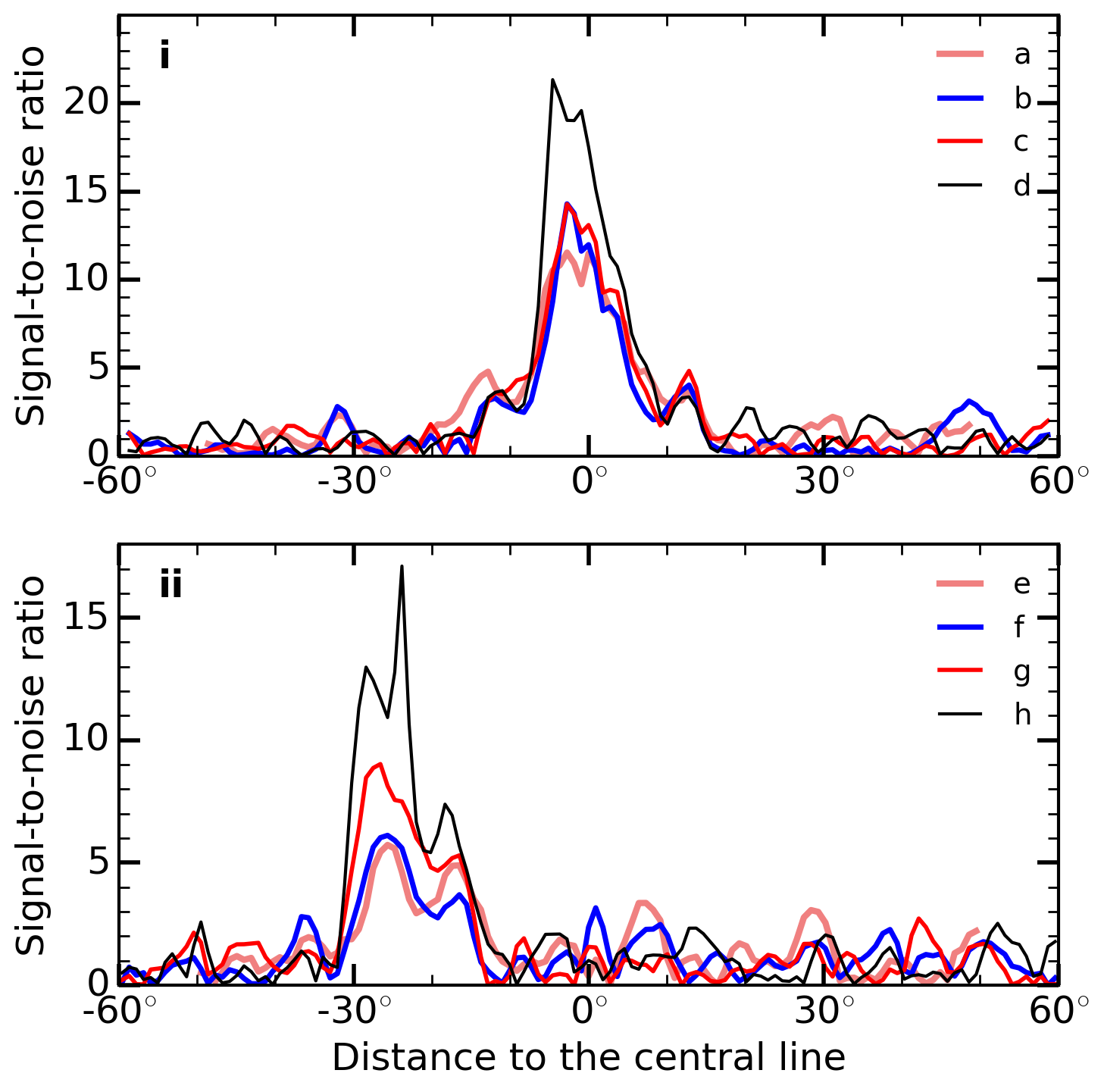}
\caption{ {Signal-to-noise ratios from Fig.~\ref{fig_snr_postel} plotted 
along horizontal cuts through the active regions  NOAA~12192 (panel i, see blue horizontal lines in Figs.~\ref{fig_snr_postel}a--d) and NOAA~12007 (panel ii, see blue horizontal lines in Figs.~\ref{fig_snr_postel}e--h).}
}\label{fig_snr_postel_slice}
\end{center}
\end{figure}

\end{document}